\documentclass[a4paper]{PoS}
\newcommand{\A}{{\mathcal{A}}}
\newcommand{\MSbar}{\overline{\rm MS}}
\def\be{\begin{equation}}
\def\ee{\end{equation}}
\def\bea{\begin{eqnarray}}
\def\eea{\end{eqnarray}}
\def\bes{\begin{subequations}}
\def\ees{\end{subequations}}
\def\bean{\begin{eqnarray*}}
\def\eean{\end{eqnarray*}}

\title{Nearly perturbative QCD coupling with lattice-motivated zero IR limit}

\ShortTitle{Lattice-motivated QCD coupling}

\author{\speaker{Gorazd Cveti\v{c}}\\
  Department of Physics, Universidad T{\'e}cnica Federico Santa Mar{\'\i}a, Casilla 110-V, Valpara{\'\i}so, Chile\\
          E-mail: \email{gorazd.cvetic@usm.cl}}

\abstract{The product of the gluon dressing function and the square of the ghost dressing function in the Landau gauge can be regarded to represent, apart from the inverse power corrections $1/Q^{2 n}$, a nonperturbative generalization ${\mathcal{A}}(Q^2)$ of the perturbative QCD running coupling $a(Q^2 )$ ($\equiv \alpha_s(Q^2)/\pi$). Recent large volume lattice calculations for these dressing functions strongly indicate that such a generalized coupling goes to zero as ${\mathcal{A}}(Q^2) \sim Q^2$ when the squared momenta $Q^2$ go to zero ($Q^2\ll 1 \ {\rm GeV}^2$). We construct such a QCD coupling $\A(Q^2)$ which fulfills also various other physically motivated conditions. At high momenta it becomes the underlying perturbative coupling $a(Q^2)$ to a very high precision. And at intermediately low momenta $Q^2 \sim 1 \ {\rm GeV}^2$ it gives results consistent with the data of the semihadronic $\tau$ lepton decays as measured by OPAL and ALEPH. The coupling is constructed in a dispersive way, ensuring as a byproduct the holomorphic behavior of $\A(Q^2)$ in the complex $Q^2$-plane which reflects the holomorphic behavior of the spacelike QCD observables. Application of the Borel sum rules to $\tau$-decay V + A spectral functions allows us to obtain values for the gluon (dimension-4) condensate and the dimension-6 condensate, which reproduce the measured OPAL and ALEPH data to a significantly better precision than the perturbative $\MSbar$ coupling (+OPE) approach. The comparison with the experimental V-channel Adler function, related with the $e^+ e^− \to$ hadrons ratio, at low $Q^2 \sim 1 \ {\rm GeV}^2$, also gives results considerably better than with the usual $\MSbar$ pQCD+OPE approach.}

\FullConference{The European Physical Society Conference on High Energy Physics\\
		5-12 July, 2017\\
		Venice}

\begin{document}

\section{Motivation}
In perturbative QCD (pQCD), the running coupling $a(Q^2) \equiv \alpha_s(Q^2)/\pi$ as a function of $Q^2 \equiv - q^2$ has Landau singularities, i.e., singularities outside the negative semiaxis in the complex $Q^2$-plane, in most renormalization schemes, including $\MSbar$ and related schemes. On the other hands, the general principles of quantum field theories, namely locality, microcausality and unitarity, imply that the spacelike physical quantities ${\cal D}(Q^2)$, such as current correlators and hadronic structure functions, are holomorphic (analytic) functions of $Q^2$ in the entire complex plane with the exception of the negative semiaxis, $Q^2 \in \mathbb{C} \backslash (-\infty, -M_{\rm thr}^2]$, where $M_{\rm thr.}$ is a threshold scale of the order of the light meson mass. Such quantities are evaluated as functions of the QCD running coupling $a(\kappa Q^2)$ where $\kappa \sim 1$ is a chosen renormalization scale parameter, ${\cal D}(Q^2)_{\rm ev.} = {\cal F}(a(\kappa Q^2))$. Since $a(\kappa Q^2)$, due to the mentioned Landau singularities, does not share the holomorphic properties that ${\cal D}(Q^2)$ must have, the evaluated expressions  ${\cal D}(Q^2)_{\rm ev.}$ (such as truncated perturbation series for the leading-twist and higher-twist terms) have wrong holomorphic properties. Furthermore, at low $|Q^2|$, due to the vicinity of the Landau singularities the evaluation of $a(\kappa Q^2)$ and thus of ${\cal D}(Q^2)_{\rm ev.}$ becomes unreliable. For these reasons, it is preferrable to evaluate ${\cal D}(Q^2)$ using a holomorphic analog $\A(Q^2)$ of the pQCD coupling $a(Q^2)$, i.e., $\A(Q^2)$ based on $a(Q^2)$, but in contrast to it, has no Landau singularities, i.e., $\A(Q^2)$ is a holomorphic function for $Q^2 \in \mathbb{C} \backslash (-\infty, -M_{\rm thr}^2]$.

A first version of such a coupling, called Analytic Perturbation Theory (APT), was constructed in \cite{ShS}. $\A^{\rm (APT)}(Q^2)$ has the same discontinuity along the negative axis $Q^2 = - \sigma < 0$ as the underlying pQCD coupling $a(Q^2)$, but the Landau discontinuities and singularities of $a(Q^2)$ along the positive axis are eliminated in the dispersive integral representation of the coupling. Later, several other couplings were constructed with the dispersive approach, where the discontinuity at low $\sigma >0$ was changed or parametrized so that the coupling fulfilled certian physically-motivated conditions \cite{otherA,mes2,CV12,NestBook,3d}. These coupling are IR-finite, $\A(0) < \infty$. Analytization of the pQCD beta function $\beta(a)/a$ also leads to a holomorphic coupling $\A(Q^2)$ \cite{Nest1}, but it is infinite at $Q^2=0$. Light-front holography approach to QCD \cite{Brod} gives an IR-finite coupling $\A(Q^2)_{\rm LFH} \propto \exp(-Q^2/Q_0^2)$ where $Q_0 \sim 1$ GeV.   

The general algorithm for the construction of the higher power analogs $a(Q^2)^n \mapsto \A_n(Q^2)$ ($\not= \A(Q^2)^n$ in general) in such holomorphic frameworks was presented in \cite{CV12} (integer $n$) and in \cite{GCAK} ($n$ noninteger).

The mentioned dispersive approaches can also be applied directly to (spacelike) physical quantities ${\cal D}(Q^2)$ to enforce the correct holomorphic and physical properties, cf.~\cite{DM,mes2,NestBook}. We will not pursue this line here.

\section{Construction of $\A(Q^2)$}

Here we will describe briefly the construction of the coupling $\A(Q^2)$ of Refs.~\cite{3d}. Having the (underlying) pQCD coupling $a(Q^2)$, in a given renormalization scheme, we will impose the following physically-motivated requirements on the coupling $\A(Q^2)$:

\noindent
1. $\A(Q^2)$ is a holomorphic function for $Q^2 \in \mathbb{C} \backslash (-\infty, -M_{\rm thr}^2]$. 

\noindent
2. At high $|Q^2| \gg 1 \ {\rm GeV}^2$ we have practically equality $\A(Q^2)=a (Q^2)$ (pQCD at high $|Q^2|$).

\noindent
3. At intermediate $|Q^2| \sim 1 \ {\rm GeV}^2$, the $\A(Q^2)$-approach reproduces the well measured semihadronic $\tau$-decay physics.

\noindent
4. At low $|Q^2| \lesssim 0.1 \ {\rm GeV}^2$, we have $\A(Q^2) \sim Q^2$, as suggested by lattice results for the Landau gauge gluon and ghost propagators \cite{LattcoupNf}.

It turns out that the property 1 will be a byproduct of the construction of $\A(Q^2)$ by the above properties $2$-$4$.

\vspace{0.4cm}

First, we will explain the property 4. We recall that in pQCD we have for $a (Q^2) \equiv \alpha_s(Q^2)/\pi$ 
\be
a (Q^2)  =  a (\Lambda^2) Z_{\rm gl}^{(\Lambda)}(Q^2) Z_{\rm gh}^{(\Lambda)}(Q^2)^2/Z_1 ^{(\Lambda)}(Q^2)^2,
\label{alatt}
\ee  
where $Z_{\rm gl}$, $Z_{\rm gh}$, $Z_1$ are the dressing functions of the gluon and ghost propagator, and of the gluon-ghost-ghost vertex.
In the Landau gauge, $Z_1^{(\Lambda)}(Q^2)=1$ to all orders \cite{Tayloretal}. Hence
\be
\A_{\rm latt.}(Q^2)  \equiv  \A_{\rm latt.}(\Lambda^2) Z_{\rm gl}^{(\Lambda)}(Q^2) Z_{\rm gh}^{(\Lambda)}(Q^2)^2 \ .
\label{Alatt}
\ee
\be
\A_{\rm latt.}(Q^2) =  \A(Q^2) + \Delta \A_{\rm NP}(Q^2) \ .
\label{AlattA}
\ee
Since $\A_{\rm latt.}(Q^2) \sim Q^2$ when $Q^2 \to 0$, no finetuning at $Q^2 \to 0$ implies
\be 
\Delta \A_{\rm NP}(Q^2) \sim Q^2 \qquad {\rm and} \quad \A(Q^2) \sim Q^2
\qquad (Q^2 \to 0)
\label{noft}
\ee
The coupling $\A(Q^2)$ thus also goes to zero when $Q^2 \to 0$, this is the mentioned property 4.

\vspace{0.4cm}

\begin{figure}[htb] 
\centering\includegraphics[width=105mm,height=60mm]{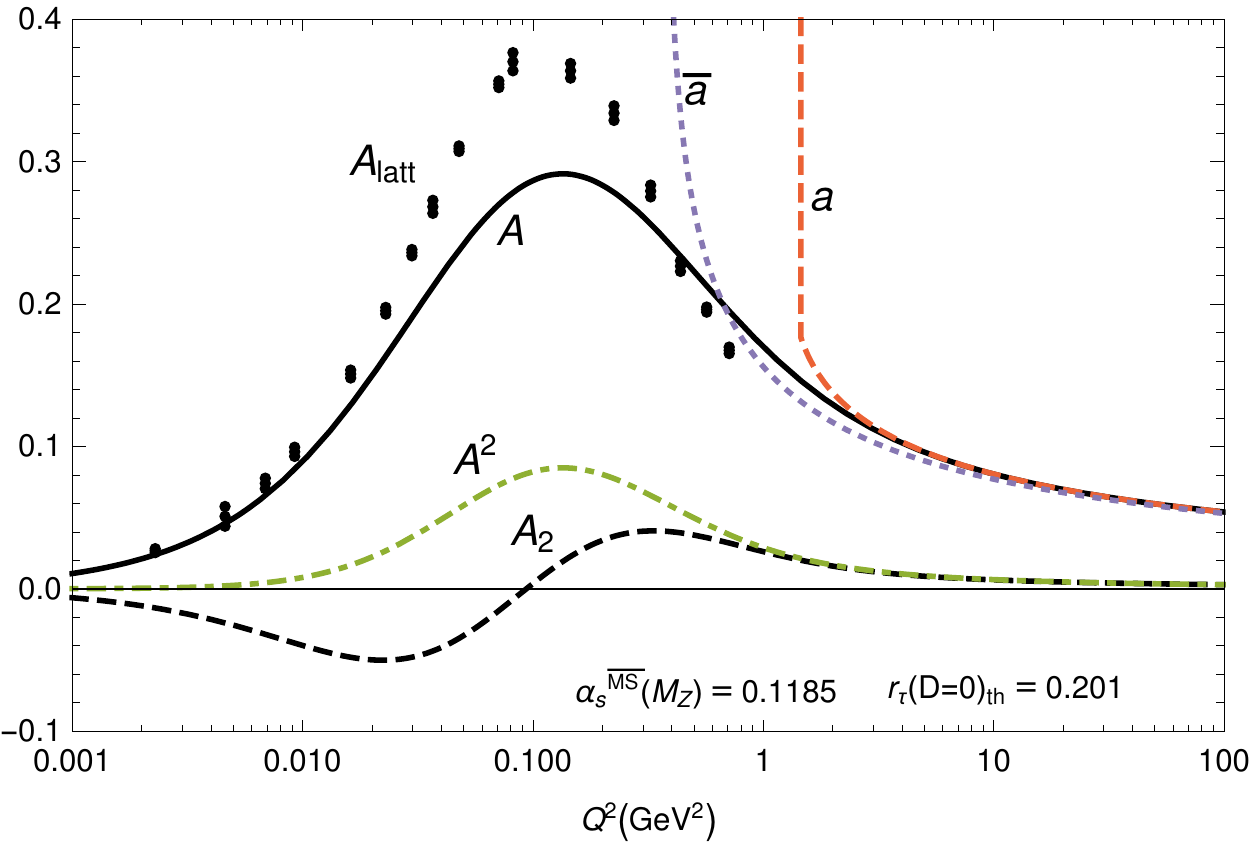}   
\vspace{-0.4cm}
\caption{\footnotesize  Coupling $\A$ at positive $Q^2$ (solid curve) and the underlying pQCD coupling $a$ (light dashed curve), both in 4-loop MiniMOM scheme. Included is $\A_2$ (dashed curve) which is the $\A$-analog of power $a^2$, and the naive (i.e., unusable) power $\A^2$ (dot-dashed curve). The usual $\MSbar$ scheme pQCD coupling ${\overline a}$ (dotted curve) is also included. The lattice coupling $\A_{\rm latt.}$ (Ref.~\cite{LattcoupNf}, first entry), with $Q^2$ rescaled as explained in the text, is presented as points with bars.}
\label{FigAa}
 \end{figure}

Now we will construct $\A(Q^2)$ such that the mentioned properties 2 and 4 can be enforced. The dispersive relation for $a (Q^2)$ is
\begin{equation}
a (Q^2) = \frac{1}{\pi} \int_{\sigma= - {Q^2_{\rm br}} - \eta}^{\infty}
\frac{d \sigma {\rho}_a (\sigma) }{(\sigma + Q^2)}
   \qquad (\eta \to +0),
\label{adisp}
\end{equation}
where $Q^2=Q^2_{\rm br} > 0$ is the branching point for Landau singularities, and $\rho_a (\sigma) \equiv {\rm Im} \; a (Q^2=-\sigma - i \epsilon)$ is the discontinuity (spectral) function of $a$. The dispersive relation for the corresponding $\A(Q^2)$ is
\be
\A(Q^2) = \frac{1}{\pi} \int_{\sigma=M^2_{\rm thr}-\eta}^{\infty} \frac{d \sigma \rho_{\A}(\sigma)}{(\sigma + Q^2)} 
\qquad (\eta \to +0),
\label{Adisp}
\ee
where $\rho_{\A}(\sigma) \equiv {\rm Im} \; \A(Q^2=-\sigma - i \varepsilon)$. At high positive $\sigma_0 > M_0$ ($\sim 1$ GeV) we expect $\rho_{\A} = \rho_a$, but at low positive $\sigma < M_0^2$ we expect $\rho_{\A} \not= \rho_a$. Here, $M_0^2$ is a pQCD onset-scale. The a priori unknown behavior of $\rho_{\A}$ in the low-$\sigma$ regime ($\sigma < M_0^2$) will be parametrized with several delta functions (peaks), specifically three delta functions. This means
\bea
\rho_{\A}(\sigma) &=&  \pi \sum_{j=1}^{3} {\cal F}_j \; \delta(\sigma - M_j^2)  + \Theta(\sigma - M_0^2) \rho_a (\sigma) \ .
\label{rhoA}
\\
\Rightarrow \;\;
\A(Q^2) &=&   \sum_{j=1}^3 \frac{{\cal F}_j}{(Q^2 + M_j^2)} + \frac{1}{\pi} \int_{M_0^2}^{\infty} d \sigma \frac{ \rho_a (\sigma) }{(Q^2 + \sigma)} \ .
\label{AQ2}
\eea
At $|Q^2| > 1 \ {\rm GeV}^2$, $\A(Q^2)$ should practically coincide with pQCD (property 2), so we require 
\be
\A(Q^2) - a (Q^2) \sim \left( \frac{{\Lambda_L}^2}{Q^2} \right)^5 \quad
(|Q^2| > {\Lambda_L}^2 \sim 0.1-1 \ {\rm GeV}^2 ) \ .
\label{Aadiff2}
\ee
This (property 2), and the lattice condition $\A(Q^2) \sim Q^2$ at $Q^2 \to 0$ (property 4), give 5 conditions
\bea
- \frac{1}{\pi} \int_{M_0^2}^{\infty} d \sigma \frac{\rho_a (\sigma)}{\sigma} &=& 
\sum_{j=1}^3  \frac{{\cal F}_j}{M_j^2} \ ;
\label{Q2to0}
\\
\frac{1}{\pi} \int_{-Q_{\rm br}^2}^{M_0^2} d \sigma \sigma^k \rho_a (\sigma) &=& \sum_{j=1}^3 {\cal F}_j M_j^{2 k}  \quad (k=0,1,2,3) \ .
\label{1u}
\eea
But we have 7 parameters, we need 7 conditions, i.e.,  two more: 

\noindent
a) $Q^2_{\rm max} \approx 0.135 \ {\rm GeV}^2$ by lattice calculations, where $\A(Q^2_{\rm max}) = \A_{\rm max}$ (extension of property 4).

\noindent
b) $\A$-coupling framework should reproduce the correct value $r^{(D=0)}_{\tau} \approx 0.20$ (cf.~\cite{ALEPH2}) of the (QCD-part of the) ratio of the semihadronic $\tau$ decay width (property 3), where
\be
r^{(D=0)}_{\tau, {\rm th}} = \frac{1}{2 \pi} \int_{-\pi}^{+ \pi}
d \phi \ (1 + e^{i \phi})^3 (1 - e^{i \phi}) \
d(Q^2=m_{\tau}^2 e^{i \phi};D=0) \ .
\label{rtaucont}
\ee
Here, $d(Q^2;D=0)$  is the massless Adler function, $d(Q^2;D=0) = -1 - 2 \pi^2 d \Pi(Q^2; D=0)/d \ln Q^2$ ($\Pi$ is the vector or axial current correlator), and its perturbation expansion is known up to $\sim a^4$. 
In our approach, $a (Q^2)^n \mapsto \A_n(Q^2)$ ($\not= \A(Q^2)^n$) \cite{CV12}
\be
 d(Q^2;D=0)_{\rm an}^{[4]} =  \A(Q^2) + d_1 \A_{2}(Q^2) +  d_2 \A_{3}(Q^2) +  d_3 \A_{4}(Q^2).
\label{dan}
\ee
These 7 conditions  (with $r_{\tau, {\rm th}}^{(D=0)}=0.201$) then give the values of the parameters of the coupling (cf.~Ref.~\cite{3d} 2nd entry)
\bean
M_0^2 =  8.719  \ {\rm GeV}^2; \quad M_1^2 = 0.053 \ {\rm GeV}^2, \quad M_2^2&=&0.247  \ {\rm GeV}^2, \quad M_3^2 = 6.341  \ {\rm GeV}^2;
\label{M2}
\\
{\cal F}_1  =  -0.0383 \ {\rm GeV}^2, \quad {\cal F}_2 &=& 0.1578 \ {\rm GeV}^2, \quad {\cal F}_3 = 0.0703 \ {\rm GeV}^2.
\label{calFj}
\eean
The underlying ($N_f=3$) pQCD coupling $a$ was constructed in the (4-loop) lattice MiniMOM scheme \cite{MiniMOM}, because the lattice results  \cite{LattcoupNf} for $\A_{\rm latt.}(Q^2)$ (at low positive $Q^2$) were obtained in this scheme. We do, however, rescale $Q^2$ from the MiniMOM ($\Lambda_{\rm MM}$) to the usual $\Lambda_{\MSbar}$-scale convention. The results are presented in Fig.~\ref{FigAa}. We note that all the locations of the delta functions, $\sigma_j=M_j^2$ ($j=1,2,3$), and $M_0^2$, turned out to be positive, i.e., the coupling $\A(Q^2)$ has no cut along the positive axis and is thus, as a consequence, holomorphic. The threshold mass is $M_{\rm thr} = M_1 \approx 0.23$ GeV, which is, as expected, in the regime of the light meson masses.

 \section{Conclusions}

 A QCD coupling $\A(Q^2)$ was constructed, in the lattice MiniMOM scheme, rescaled to the usual $\Lambda_{\MSbar}$-scale convention. It has the following properties:

\noindent
A)  $\A(Q^2)$ reproduces the pQCD results at high momenta $|Q^2| > 1 \ {\rm GeV}^2$.

\noindent
B) $\A(Q^2) \sim Q^2$ at low momenta $|Q^2| \lesssim 0.1 \ {\rm GeV}^2$, as suggested by high-volume lattice results.

\noindent
C) $\A(Q^2)$ at intermediate momenta $|Q^2| \sim 1 \ {\rm GeV}^2$ reproduces the well the measured physics of the inclusive semihadronic $\tau$-lepton decay.

\noindent
D) $\A(Q^2)$, as a byproduct of construction, possesses the attractive holomorphic behavior shared by QCD spacelike physical quantities.

The usual $\MSbar$ pQCD coupling $a (Q^2;\MSbar) \equiv \alpha_s(Q^2;\MSbar)/\pi$ shares with the coupling $\A$ only the property A (high-momentum), but on the other three properties it is either worse (point C) or it fails (points B and D).

We applied the Borel sum rules to $\tau$-decay V + A spectral functions, and we obtained values for the gluon (dimension-4) condensate and the dimension-6 condensate by fitting to the measured OPAL and ALEPH data. The fitting turned out to be significantly better than with the perturbative $\MSbar$ coupling (+OPE) approach. Further, when we compared the obtained theoretical and the experimental V-channel Adler function ${\cal D}_V(Q^2)$, related with the $e^+ e^− \to$ hadrons ratio, at low $Q^2 \sim 1 \ {\rm GeV}^2$, the results were considerably better than with the usual $\MSbar$ pQCD+OPE approach. We refer for details to Refs.~\cite{3d} (in the 3-loop and 4-loop MiniMOM, respectively).

\end{document}